%% file: main.tex
\documentclass[twocolumn,twoside]{IEEEtran}

\usepackage{etoolbox}
\usepackage{epsfig}
\usepackage{array}
\usepackage{url}
\usepackage{tikz}
\usetikzlibrary{shadows}
\usepackage{ifthen}
\usepackage{color}
\usepackage{pgfplots}
\pgfplotsset{compat=newest}
\usepackage{amsmath,amssymb}
\usepackage{dblfloatfix}
\newcommand{\code}[1]{{\small \tt #1}}
\newcommand{\de}[1]{{\it #1}}
\begin{document}

\author{Alex Biryukov\thanks{University of Luxembourg, E-mail: alex.biryukov@uni.lu},
Ivan Pustogarov\thanks{University of Luxembourg, E-mail: ivan.pustogarov@uni.lu}, 
Fabrice Thill\thanks{University of Luxembourg, E-mail: fabrice.thill.001@student.uni.lu},
Ralf-Philipp Weinmann\thanks{University of Luxembourg, E-mail: ralf-philipp.weinmann@uni.lu}
}

\title{Content and popularity analysis of\\ Tor hidden services}


\maketitle 
\begin{abstract}
Tor hidden services allow running Internet services while protecting
the location of the servers. Their main purpose is to enable freedom 
of speech even in situations in which powerful adversaries try to suppress 
it. However, providing location privacy and client anonymity also
makes Tor hidden services an attractive platform for every kind of 
imaginable shady service. The ease with which Tor hidden services can be set up
has spurred a huge growth of anonymously provided Internet services of both types.
In this paper we analyse the landscape of Tor hidden services.
We have studied 39824 hidden service descriptors collected
on 4th of Feb 2013: we scanned them for open
ports; in the case of 3050 HTTP services, we analysed and classified
their content. We also estimated the popularity
of hidden services by looking at the request rate for hidden service
descriptors by clients. We found that while the content of Tor hidden
services is rather varied, the most popular hidden services are related to botnets.
We also propose a method for opportunistic deanonymisation of
Tor Hidden Service clients. In addtiton, we identify past attempts
to track ``Silkroad'' by consensus history analysis.

\end{abstract}

\input{intro}
\input{background}
\input{portscanning}

\input{contanalysis}
\input{popularity}

\input{track-clients}
\input{silkanalysis}

\input{conclusion}
\bibliographystyle{IEEEtran}
\bibliography{main}

\end{document}

%% file: intro.tex
\section{Introduction}
\label{sec:introduction}

Tor hidden services allow Tor users to offer various Internet services like
web publishing or messaging while keeping the location of said
services hidden. Other Tor users can connect to them through so-called
\de{rendezvous points}. Since Tor added support for hidden services in 2004, many of
them have emerged; some enable freedom of speech (New Yorker's Strongbox \cite{strongbox},
Wikileaks \cite{wikileaks}) while others allow for the exchange of contraband (the Silk Road
market place \cite{silkroad}) or are used by botnets (Skynet \cite{rapid7skynet}) for
hiding the location of command and control centers. More mundane services such as the
DuckDuckGo search engine \cite{duckduckgo} also exist as Tor hidden services.

In order to find a hidden service one can use a Hidden Wiki \cite{hiddenwiki}
or one of the specialized search engines~\cite{ahmia}. 
Though such resources are of a great use, the number of onion addresses they 
offer is quite limited\footnote{At the time of writing, three Hidden Wikis linked
to 624 onion addresses and \code{ahmia.fi} had 1033 onion addresses over which the search
could be made.} and is not sufficient for the full analysis of the landscape of Tor hidden services.
At the same time mining the full collection of onion addresses is prevented by several objective reasons:
firstly, many hidden service operators are not interested in their existence to be widely known.
Secondly, Tor is designed in such a way that no central entity stores the entire list of onion addresses.
Furthermore, hidden services only rarely link to each other -- which impedes traditional crawling.

In this paper we first present a study of deployed Tor hidden services.
In order to overcome the above mentioned limitations, we exploited a flaw which was
discovered in \cite{SnPSybil} and was present in Tor versions before 0.2.4.10-alpha.
Using this approach we collected 39824 unique onion addresses
on February 4rd, 2013 by running 58 Amazon EC2 instances.
We scanned the obtained list of hidden services for open ports, estimated their popularity,
and classified their content. 
Second, we propose a method for opportunistic deanonymisation of
Tor Hidden Services clients. Third, we make a statistical analysis of the Tor
consensus history in order to identify past attemps to track ``Silkroad''
hidden service.

%% file: background.tex
\section{Background}
\label{sec:background}
In this sections we briefly describe Tor hidden services architecture and
mention key points of the flaws which allowed us to mine the collection
of onion addresses and opportunistically deanonymise Tor hidden service clients.
For a more detailed description we refer the reader to \cite{SnPSybil}.

\subsection{Tor hidden services}

Tor is a low-latency anonymity network based on the ideas of onion
routing and telescoping. Clients can have anonymous communication to a
server by proxying their traffic through a chain ({\it Circuit}) of three Tor
relays. Tor is a volunteer-based network and anybody can run and
advertise its own Tor relay. However there is a small number of so called
Tor authorities which monitor self-advertised relays and maintain and
distribute among clients a list ({\it Consensus}) of relays which satisfy
some conditions: a relay should be be up; the number of Tor relays
per one IP address should not exceed two.

Tor {\it Hidden Services} are a feature which was introduced in 2004
to add responder anonymity to Tor. Specifically, hidden services allow
running an Internet service (e.g. a Web site,
SSH server, etc.) so that the clients of the service do not know its
actual IP address. This is achieved by routing all communication between
the client and the hidden service through a {\em rendezvous point} which
connects anonymous circuits from the client and the server.

In order to make an Internet service available as a Tor hidden
service, the operator (Bob) configures his Tor software which automatically
generates new RSA key pair.
The base-32 encoding \textless z\textgreater{} of the first 10 bytes of SHA-1 digest of the
key becomes the identifier of the hidden service (its onion address is
constructed from \textless z\textgreater{} by adding ".onion" part).

When a client (Alice) wants to communicate with the hidden service,
she needs to know not only its onion address, but also the public key
and the list of introduction points. Due to this Bob generates two 
service descriptors which contain this information
and uploads them to 6 hidden service directories. A hidden services directory
is a Tor relay which has an HSDir flag assigned by Tor authorities;
a Tor relay needs to be operational
for at least 25 hours to obtain this flag. The choice of hidden service
directories deterministically depends on the onion address and current time. A hidden service
changes its responsible hidden service directories every 24 hours and
HS directories responsible for the previous timeperiod erase its descriptor
from the memory\footnote{Note
that the expression to compute next responsible HS directories is deterministic
and an attacker can easily inject relays to the Tor network which would
become responsible HS directories.}.
This results in that the information about all existing hidden services
at given time is distributed over many Tor relays.
If an adversary wants to collect all existing onion addresses she would
need to either own quite many\footnote{At the time of the experiments an attacker
would need to own more than 300 IP addresses for at least 27 hours.} IP addresses or run a small
set of relays from small number of IP addresses for significant amount of time.
This is enforced by that  the maximum number of Tor relays
on a single IP address that Tor authorities include to the Consensus is 2.
If more than two relays are running on the same IP address, only two relays with
the highest-most measured bandwidth will appear in the Consensus.
This prevents an attacker from injecting many Tor relays from a single IP address
in order to collect onion addresses.

A flaw discovered in \cite{SnPSybil} allowed us to overcome the limitation of
two Tor relays per IP address and use only 58 IP addresses to gather our
collection of onion addresses.
We now describe the key points of the flaw. While only two relays per IP
appear in the Consensus, all running relays
are monitored by the authorities; more importantly, statistics on them
is collected, including the uptime which is used to decide which
flags a relay will be assigned.
We call relays appearing in the Consensus {\em active relays} and
those which run at the same IP address but do not
appear in the Consensus {\it shadow relays}. Whenever one of the
active relays becomes unreachable and disappears from the Consensus,
one of the shadow relays becomes active, i.e. appears in
the Consensus. This new active relay will have all the
flags corresponding to its real run time and not to the time for which
it was in the consensus. We call this technique {\it shadowing}.

An attacker can use this as follows. She can rent $n$ IP addresses and run
$m$ relays on each of them for 25 hours thus running $n \times m$ Tor instances in
total. Though only $2n$ of them should appear in the consensus, 
at the end of 25 hour time period all $n \times m$ relays will have HSDir flags.
The idea is to gradually make active relays unreachable to the Tor authorities
so that shadow relays become active and thus gradually become hidden service
directories for all hidden services during 24 hours, thus collecting hidden services'
public keys (from which onion addresses are easily derived) and measuring clients
request rates.

\subsection{Guard nodes}
In order to significantly reduce the probability of traffic confirmation attacks
\cite{Ling}, \cite{Serj} Tor developers introduced the concept of \emph{entry guard nodes}.
A Tor client initially selects a set of three guard nodes from among Tor relays which have
a Guard flag assigned to them. Whenever less than two guard nodes
from the set are reachable, new guard nodes are chosen.
A guard node remains in the set for a random duration between 30 and 60 days.
Then it is marked as expired and removed from the set.
Whenever a circuit is established, one node from the set of Guard nodes is
used for the first hop.
An attack presented in \cite{SnPSybil} allowed an attacker to deanonymise the operator
of a hidden service once an attacker's relay becomes a Guard node of the hidden service.
In order to mount this attack, the responsible hidden service directory controlled
by the attacker sends a specific traffic signature to the hidden service immediately after
the hidden service uploads its descriptor. This signature is then detected at the guard node.
In this paper we adapt this attack to deanonymise clients of a Tor hidden service.

%% file: portscanning.tex
\section{Port scanning hidden services}
\label{sec:port-scanning}
In this section we provide statistics on open ports of Tor hidden
services.  We scanned the full collection of 39824 onion addresses at
different times between 14 and 21 Feb 2013.  At the time of the scans
hidden service descriptors were available for 24511 addresses. In
total, 22007 ports were found open on these. For hidden services for
which descriptors were available, we obtained a coverage of 87\% of
all ports. The full coverage could not be achieved since we scanned different
port ranges on different days and in a number of cases hidden services
which we partially scanned on one day went off-line the day of the next scan;
when scanning some hidden services we were persistently getting timeout errors.

During the scan we noticed that a large amount of hidden services did
not have any open ports, however when scanned for port 55080 they
returned an error message different from the usual error message.
According to the Rapid7 blog post \cite{rapid7skynet} port 55080
corresponds to hidden services created on computers infected by a
botnet malware called ``Skynet''. The observation is explained by the
fact that the malware immediately closes any connection to this port
unless it has been set up as a connection forwarder. We received such
an ``abnormal'' error message for port number 55080 only and counted
such events as open ports. 

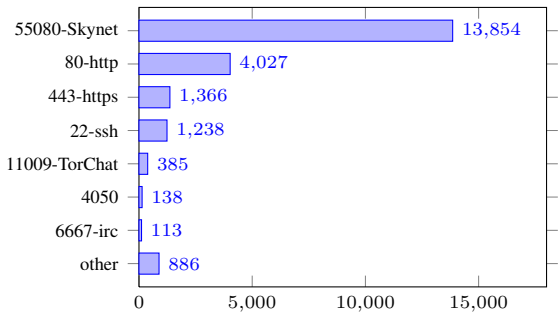
\begin{figure}[t]
\begin{tikzpicture}
\begin{axis}[scaled ticks=false,
xbar, xmin=0, xmax=18000,
width=7cm, height=5.3cm,
bar width=8pt,
symbolic y coords={other,6667-irc,4050,11009-TorChat,22-ssh,443-https,80-http,55080-Skynet},
ytick=data,
x tick label style={font=\scriptsize},
y tick label style={font=\scriptsize},
nodes near coords, nodes near coords align={horizontal},
every node near coord/.append style={font=\scriptsize}
]
\addplot coordinates {(886,other) (113,6667-irc) (138,4050)%
                      (385,11009-TorChat) (1238,22-ssh) (1366,443-https)%
		      (4027,80-http) (13854,55080-Skynet)};
\end{axis}
\end{tikzpicture}
\caption{Open ports distribution}
\label{fig:openports-dist}
\end{figure}

The open ports distribution is shown in Fig.~\ref{fig:openports-dist}.
Port number 55080 is the most frequent one, found open on more than 50\% of
all onion addresses. This can be used to estimate the number of computers
infected by ``Skynet''. HTTP and HTTPS services constitute
22\% and SSH services are run by 5\% of hidden services. Ports not shown on 
Fig.~\ref{fig:openports-dist} have counts of less than 50 and are grouped together
under ``Other'' label. In total we found 495 unique port numbers.

During our port scan we discovered that a number of hidden services provided HTTPS
access. We discovered that in 1,225 cases the certificates were self signed and
the certificates' common names did not match the requested host names.
In 1,168 cases the certificate common name was ``esjqyk2khizsy43i.onion'' which
is hosted at free onion hosting service ``TorHost''. We found 34
hidden services using certificates containing common names
corresponding to their public DNS names, allowing for deanonymization of the service.

%% file: contanalysis.tex
\section{Content analysis}
\label{sec:contanalysis}
In this section we present an analysis of the content of hidden services
which provide HTTP(S) access: we classify them both according to
topics and languages.
We excluded port 55080 and tried to connect to the remaining 8,153
destinations (onion address:port pairs) using HTTP and HTTPS.  We
performed the crawl 2 months after the port scan\footnote{We excluded
all binary data such as images, executables, etc.}. At the time of the
crawl, 7,114 ports were open -- of these, we were able to connect to
6,579 using either HTTP or HTTPS.  Table \ref{tab:http-ports} shows
the number of hidden services that offered HTTP or HTTPS services.

\begin{table}[!h]
\centering
  \begin{tabular}{| c | c |}
    \hline
    \textbf{Port Num} & \textbf{\# of onion addresses} \\ \hline \hline
      80 & 3741 \\ \hline
     443 & 1289 \\ \hline
      22 & 1094 \\ \hline
    8080 & 4    \\ \hline
   Other & 451  \\ \hline
  \end{tabular}
  \caption{HTTP and HTTPS access}
  \label{tab:http-ports}
\end{table}

About half of the destinations were inappropriate for classification so
we excluded them and ended up with 3050 destinations. In more detail we excluded:
2348 destinations which contained less then 20 words of text (this included 1092
messages from port 22 which were SSH banners);
1108 destinations at port 443 which had corresponding copies at port 80;
73 destinations which returned an error message embedded in an HTML page.

For language detection we used ``Langdetect'' \cite{langdetect} software.
The vast majority of the hidden services (84\%) were in English. This is an expected result and
corresponds to the statistics for the public Internet \cite{Gerrand07}.
Overall we found hidden services in 17 different languages. Content
was offered in the following languages besides English (each constituting less than
3\%): German, Russian, Protuguess, Spanish, French, Polish, Japanese, Italian, Czech, Arabic, Dutch, Basque, Chinese, Hungarian, Bantu, Swedish.

We used the software ``Mallet'' \cite{mallet} and the web service
``uClassify'' \cite{uclassify} for automatic topic classification. We
considered only hidden services which offered pages in English (2,618
hidden services in total).  Among them, 805 hidden services showed the
default page of the Torhost.onion\footnote{torhostg5s7pa2sn.onion}
free anonymous hosting service. We classified the remaining 1,813 onion
addresses into 18 different categories.

Resources devoted to drugs, adult content, counterfeit (selling counterfeit products,
stolen credit card numbers, hacked accounts, etc.), and weapons 
constitute 44\%. 
The remaining 56\% are devoted to a number of different topics: ``Politics'' and ``Anonymity'' are 
among the most popular (9\% and 8\% correspondingly). In the ``Politics'' category, one can
find resources for reporting and discussing corruption, repressions, violations of 
human rights and freedom of speech, as well as leaked cables, and Wikileaks-like pages;
the category ``Anonymity'' includes resources devoted to discussion of anonymity from both
technical and political points of views as well as services which provide 
different anonymous services like anonymous mail or anonymous hosting. 

The category ``Services'' includes pages which offer money laundering, escrow services,
hiring a killer or a thief, etc.  In ``Games'' one  can find a chess server, lotteries, and
poker servers which accept bitcoins.
While making a preliminary analysis of the collected onion address names we noticed
that 15 of them had prefix ``silkroa'' (two ``official'' addresses of the silkroad marketplace
and the silkroad forum were among them). At least one of these
addresses is a a phishing site imitating the real Silk Road login interface.

\begin{figure}[t!]
\begin{center}
\includegraphics[scale=0.68]{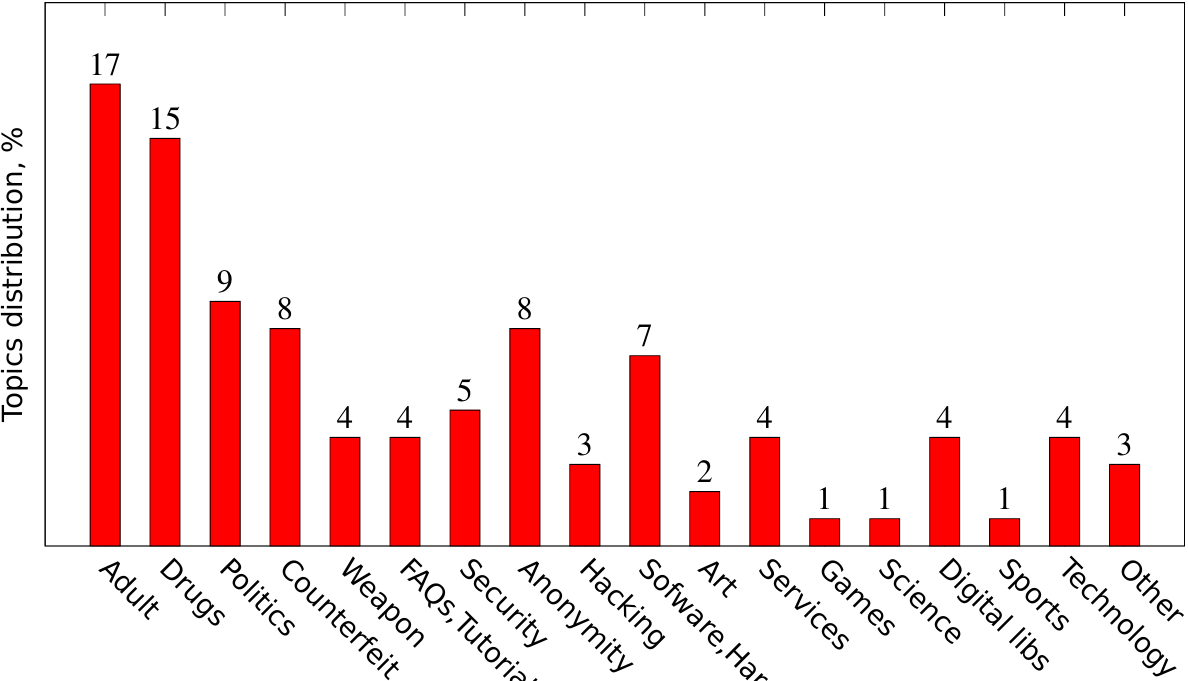}
\caption{Tor Hidden services topics distribution}
\label{fig:topic-dist}
\end{center}
\end{figure}

%% file: popularity.tex
\section{Popularity measurement}
\label{sec:popularity}
In the previous sections we have analyzed the Tor hidden services landscape
from the supply side. The analysis however will not be
complete without estimating the popularity of different hidden services among clients. 
This becomes possible since the method we used to collect onion addresses 
also allows us to get the number of client requests for each of them in a 2 hour period.
This can serve as an approximation of the popularity of hidden services. 

During our experiments, we received a total of 1,031,176 requests for 29,123
unique descriptor IDs\footnote{The descriptor ID is not equivalent to the onion address. While
the onion address remains fixed, the descriptor ID changes every 24 hours and is derived
from the onion address. Specifically, it is used to fetch hidden service's public key.
See \cite{SnPSybil} for more details.}. We used
our database of onion addressees to resolve the descriptor ID requests.
For each address in the list we computed corresponding descriptor IDs for each
day between 28 January 2013 and 8 February in order to deal with possible
wrong time settings of Tor clients. We compared this list of derived
descriptor IDs with the list of client requests. In this way we resolved 
6,113 descriptor IDs to 3,140 different onion addresses.

In order to explain the small fraction of resolved descriptor IDs, we ran
several hidden service directories for a number of days. From the log files we 
could derive that 80\% of the clients' requests were for non-existent descriptors (i.e.
which were never published). Also only 10\% of published descriptors were ever
requested by clients.
Given that the number of collected onion addresses is 39824, we believe that the small
number of resolved descriptor IDs was caused by clients requesting descriptors which
did not exist. We do not have a good explanation for this phenomenon 
(one explanation could be that specialized Hidden Service search engines
were trying to connect to services from their databases which did not
exist anymore), but this was a consistent behaviour over several months.

\newcounter{lrank}
\setcounter{lrank}{0}
\newcommand{\leftrank}{\addtocounter{lrank}{1}\arabic{lrank}}

\newcounter{rrank}
\setcounter{rrank}{21}
\newcommand{\rightrank}{\addtocounter{rrank}{1}\arabic{rrank}}

\begin{table*}[h!]
\centering
\begin{tabular}{|c|c|>{\tt}c|c||c|c|>{\tt}c|c|}
\hline
\# &RQSTS&{\rm Addr}&Desc&\#&RQSTS&{\rm Addr}&Desc\\ \hline
\leftrank & 13714 & uecbcfgfofuwkcrd.onion & Goldnet & \rightrank & 899    & qdzjxwujdtxrjkrz.onion &  Skynet \\ \hline
\leftrank & 11582 & arloppepzch53w3i.onion & Goldnet & \rightrank & 898    & 6tkpktox73usm5vq.onion &  Skynet \\ \hline
\leftrank & 11315 & pomyeasfnmtn544p.onion & Goldnet & \rightrank & 889    & kk2wajy64oip2***.onion &  Adult \\ \hline
\leftrank & 7324  & lqqciuwa5yzxewc3.onion & Goldnet & \rightrank & 781    & gpt2u5hhaqvmnwhr.onion &  Skynet \\ \hline
\leftrank & 7183  & eqlbyxrpd2wdjeig.onion & Goldnet & \rightrank & 746    & smouse2lbzrgeof4.onion &  $<$n/a$>$ \\ \hline
\leftrank & 6852  & onhiimfoqy4acjv4.onion & $<$n/a$>$   & \rightrank & 694    & xqz3u5drneuzhaeo.onion &  FreedomHosting \\ \hline
\leftrank & 6528  & saxtca3ktuhcyqx3.onion & Goldnet & \rightrank   & 667  & f2ylgv2jochpzm4c.onion &  Skynet \\ \hline
\leftrank & 4941  & qxc7mc24mj7m4e2o.onion & $<$n/a$>$   & \rightrank & 585    & kdq2y44aaas2a***.onion &  Adult \\ \hline
\leftrank & 3746  & mwjjmmahc4cjjlqp.onion & BcMine  & \rightrank & 542    & 4pms4sejqrryc***.onion &  Adult \\ \hline
\leftrank & 3678  & mepogl2rljvj374e.onion & Skynet  & \dots & \dots & \dots & \dots\\ \hline
\leftrank & 2573  & m3hjrfh4hlqc6***.onion & Adult   & 34     & 453    & dkn255hz262ypmii.onion &  SilkRoad(wiki) \\ \hline
\leftrank & 1950  & ua4ttfm47jt32igm.onion & Skynet  & \dots & \dots & \dots & \dots\\ \hline
\leftrank & 1863  & opva2pilsncvt***.onion & Adult   & 47     & 255    & dppmfxaacucguzpc.onion  &  TorDir \\ \hline
\leftrank & 1665  & nbo32el47o5cl***.onion & Adult   & \dots & \dots & \dots & \dots\\ \hline
\leftrank & 1631  & firelol5skg6e***.onion & Adult   & 62     & 172    & 5onwnspjvuk7cwvk.onion  &  BlckMrktReloaded\\ \hline
\leftrank & 1481  & niazgxzlrbpevgvq.onion & Skynet  & \dots & \dots & \dots &  \dots\\ \hline
\leftrank & 1326  & owbm3sjqdnndmydf.onion & Skynet  & 157    & 55     & 3g2upl4pq6kufc4m.onion &  DuckDuckGo \\ \hline
\leftrank & 1175  & silkroadvb5piz3r.onion & Silk Road& \dots & \dots & \dots &  \dots\\ \hline
\leftrank & 1094  & candy4ci6id24***.onion & Adult   & 250    & 30     & x7yxqg5v4j6yzhti.onion &  Onion Bookmarks \\ \hline
\leftrank & 1021  & x3wyzqg6cfbqrwht.onion & Skynet  & \dots & \dots & \dots &  \dots\\ \hline
\leftrank & 942   & 4njzp3wzi6leo772.onion & Skynet  & 547    & 10   & torhostg5s7pa2sn.onion &  Tor Host \\ \hline
\end{tabular}
\caption{Ranking of most popular hidden services}
\label{tab:popularity}
\end{table*}

Table~\ref{tab:popularity} shows the number of requests for the most popular hidden services.
We explored the five most popular addresses in more detail. Searching for them 
using the major search engines did not give any result -- this already seemed
quite strange for very popular hidden services. They only exposed port 80;
connecting to them at this port returned 503 Server errors. As a next step, we tried to
retrieve server-status pages, which succeeded. By analysing these pages we noticed
that traffic to these servers remained constant at about 330 KBytes/sec and
had about 10 client requests per second, almost exclusively POST
requests. Looking at other hidden services we discovered another 4 onion addresses with the very same
characteristics: they had port 80 open, they were returning 503 server errors, and
had server-status page available. They also had similar traffic and client request rates.
By looking at the uptime of the Apache server on the server-status pages we noticed that they could
be divided into two groups with exactly same uptime within each group. From
this we assumed that different hidden services lead to two physical servers.
Given a huge number of requests, we made a conclusion that these hidden services belong
to a very large botnet infrastructure (probably different from Skynet, we call it ``Goldnet'').
It is also worthwhile to notice that 10 onion addresses of ``Skynet''
were also among the most popular hidden services (residing between 10th and 28th places).

The Skynet bitcoin pooling servers are the second most popular, just after the probable botnet. However their request rate is
4 time lower. Bitcoin mining servers are followed by resources offering adult content (there were
8 such resources among the 30 most popular hidden services).
According to our results, the Silk Road market place is at 18th place with 1175 requests per 2 hours. 
Black Market Reloaded (another market for illegal goods)
is at 62th place with 172 requests. With regard to the popularity of 
other hidden services, Freedom hosting is at 27th place with 694
requests, and the  DuckDuckGo
search engine is at 157th place with 55 requests. The public bitcoin mining pools
Slush and Eligius had two and zero requests respectively.

%% file: track-clients.tex
\section{Tracking clients}
In \cite{SnPSybil}, the authors used a specific traffic signature for 
opportunistic deanonymisation of hidden services.
The technique they used can be easily modified for
opportunistic deanonymisation of Tor Hidden Services clients. 

Assume that an attacker controls a responsible HS directory\footnote{
Responsible hidden service directories of a hidden service are used to store the hidden
service's descriptor (which contains its public key) for a period 24
hours. The HS directories are chosen
among all Tor relays -- different hidden services usually have different responsible
HS directories.}
of a hidden service. Whenever it receives a descriptor request for that
hidden service, it sends it back encapsulated in a specific
traffic signature which will be then forwarded to the client via its Guard node.
With some probability, the client's Guard node is in the set of Guards
controlled by the attacker.
Whenever an attacker's Guard receives the traffic signature, it 
can immediately reveal the IP address of the client.

This attack has several important implications. Suppose that we can
categorize users on Silk Road into buyers and sellers.
Buyers visit Silk Road occasionally
while sellers visit it periodically to update their product pages and
check on orders. Thus,
a seller tends to have a specific pattern which allows his identification.
Catching even a small number of Silk Road sellers can seriously spoil Silk Road's
reputation among other sellers. 

\begin{figure}[h]
\begin{center}
\includegraphics[scale=0.23]{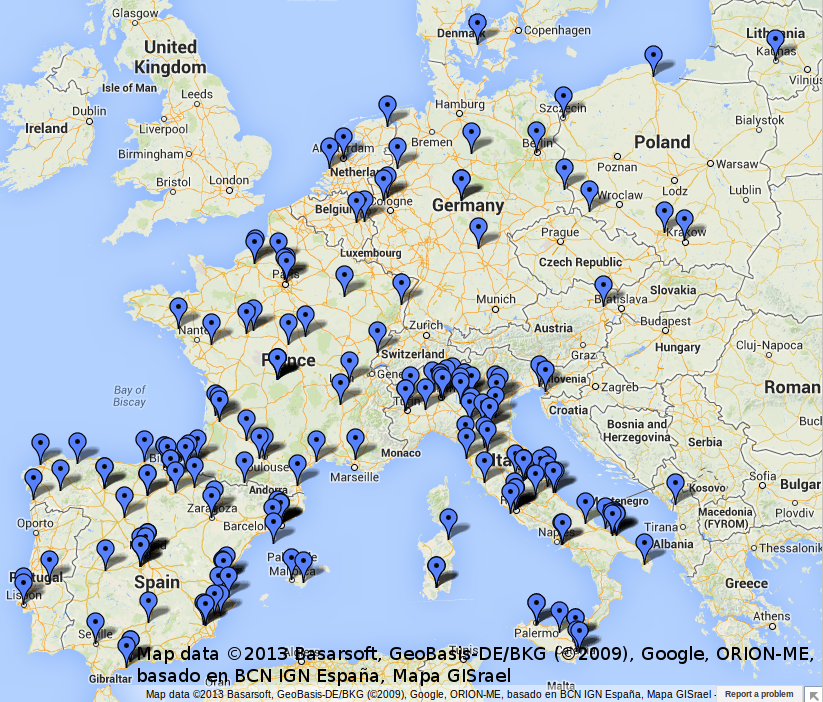}
\caption{Clients of a popular hidden service}
\label{fig:goldnetmap}
\end{center}
\end{figure}

As another application, one can collect IP addresses of clients
of a popular hidden service and compute a map
representing their geographical location. We have computed such a map for
one of the Goldnet hidden services -- in Figure \ref{fig:goldnetmap}.

%% file: silkanalysis.tex
\section{Tracking detection}
\label{sec:silkanalysis}
The current implementation of Tor allows one to control
responsible hidden service directories of a hidden service.
This enables tracking of clients requests. In this section we show
that such tracking can be identified using statistical analysis of
the consensus history. We apply our analysis to 
``Silk Road''\footnote{We only considered the
silkroadvb5piz3r.onion address} market place as an example but it can be
applied to any other service as well.

Before proceeding with the analysis we need to dwell on the details of the
Tor Hidden Services protocol. For clients to be able to connect,
a hidden service announces its existence and provides contact
information; every 24 hours it calculates two new service descriptors
and for each service descriptor 3 {\it responsible hidden service directories}
are designated from the set of Tor relays with HSDir flag.
The descriptors are then uploaded to the responsible hidden service directories.
We call an interval between two consecutive descriptor uploads as ``time period''.
In order for a relay to obtain an HSDir flag it needs to be online  for at least 25 hours. 

A relay with an HSDir flag becomes a responsible HSDir if its
fingerprint (the SHA-1 digest of its public key), is one of the 3
fingerprints that follow the hidden service descriptor's ID.
A hidden service descriptor contains all the information necessary
to establish a connection to the hidden service.
  
In our analysis we identify nodes with non-random behaviour.
First we look at the number of times a Tor relay was a responsible
directory for a hidden service. At the beginning of each time period, a
hidden service chooses 6 responsible HSDir's anew.
The probability for a relay to be chosen equals $p=6/N_{hsdir}$,
where $N_{hsdir}$ is the total number of relays with HSDir flag.
Assuming  this probability stays constant, the number of time periods
a Tor relay is chosen during $n$ days has binomial distribution
with mean $\mu=np$ and standard deviation $\sigma=\sqrt{np(1-p)}$.
If number of times a particular Tor relay was chosen 
exceeds $\mu+3\sigma$ we mark this relay as suspicious.

Second, if a Tor relay changes its fingerprint shortly before becoming a
responsible HSDir for a descriptor, it is probable that it was
done on purpose. The same is also true if a server becomes a responsible
HSDir 25 hours after its appearance in the consensus, i.e the minimum
amount of time necessary to obtain an HSDir flag. We mark a Tor relay as
suspicious if we observed such behaviour several times.

Third, we sort fingerprints (as integers) of relays with HSDir flag and calculate the ratio :$avg\_dist/distance$
where $avg\_dist$ is the average distance\footnote{By distance we mean difference where fingerprints are taken as
integer numbers} between consecutive fingerprints and $distance$ is the distance between a responsible HSDir 
and the corresponding descriptor ID. 
An attacker would try to position her relay's fingerprint as close to the two
descriptor ID as possible in order to increase her chances of staying
one of the 6 responsible HSDir's for the whole time period. 

Fourth we consider the number of times a server switched its fingerprint. Switching public keys in the 3 year period is to be expected but a high number
of switches in a short time period is unusual.

Finally we looked at servers that were responsible HSDir's for consecutive time periods.

Now we apply these rules for
the ``Silkroad'' analysis.
Silk road was launched in February 2011 and was closed by the FBI on October 2\textsuperscript{nd} 2013 \cite{silkshutdown}. 
Since we use the number of servers in our calculation of the standard deviation, and the number of HSDir more than doubled in the considered time period
\footnote{1\textsuperscript{st} of February 2011 :  757, 31\textsuperscript{st} of October 2013 :  1862}, we split our analysis in 3 years.

In the first year\footnote{1 Feb 2011 - 31 Dec 2011} there is no clear indication of tracking. A number of servers are
responsible HSDir's for many time periods and hours but (1) the time periods seem to be randomly distributed, and (2)
the servers which were responsible HSDir's more frequently than $\mu+3\sigma$ 
did not change their fingerprints nor controlled consecutive time periods.

One server shows a strange behaviour: most of the time it does not have an HSDir flag,
however in 3 occasions, it obtains it when ``Silkroad'' would choose it as a responsible HSDir.

The second year\footnote{1 Jan 2012 - 31 Dec 2012} we found our own servers which performed fingerprint changes on multiple occasions, each time with a close distance\footnote{ratio bigger than 100} between the fingerprint and the hidden service descriptor ID.

The third year \footnote{1 Jan 2013 to 31 Oct 2013} shows clear evidence of at least two cases of tracking.
From May 21\textsuperscript{st} to June 3\textsuperscript{rd} a set of servers that share the same name,
take over 1 out of 6 HSDir's for "Silk Road".  They skip only 4 time period
during this timespan. A more detailed look into these servers shows that they changed fingerprints
in order to become HSDir for "Silk Road". They are also the only responsible HSDir's that cross a ratio of 10k.
On the 31\textsuperscript{st} of August 2013 6 other Tor relays (sharing common parts in their names)
from 3 different IP addresses become the responsible HSDir's for "Silk Road".
For all 6 HSDir's the distance between their fingerprints
and respective descriptor IDs is very small.

We can conclude, that "Silk Road" has not only been tracked by us but also by other entities.
There seems to be no indication that someone tried to track "Silk Road" before November 2012
and only one entity has taken over all 6 HSDir's for a single time period, a month before the takedown by the FBI. 
Note that statistically it is impossible to distinguish attempts to track a hidden service for one
time period only from the case when a relay becomes a responsible HSDir by chance.

Based on our experiments we can conclude that 
looking for changes in fingerprints, in combination with the distance between
the descriptor ID and the fingerprint seems to be the most reliable way to detect tracking.

%% file: conclusion.tex
\section{Conclusions}
\label{sec:conclusion}
Tor hidden services are often criticized for being shelter for
resources with illegal or controversial content. The arguments used
are usually based on services such as ``Silk Road'' marketplace or
child pornography. On the contrary, Tor enthusiasts point out
that many hidden services are in fact resources devoted to human
rights, freedom of speech and information which is prohibited in
countries with oppressive regimes. Obviously both types of services
exist, but it is unclear which type prevails. Obtaining such
statistics in the past was prevented by a number of
reasons: hidden service descriptors are stored in a
distributed fashion; there is no central entity storing 
the full list of onion addresses; the number of onion
addresses published on the Internet is far from exhaustive.

In this paper we have mined a collection of hidden services
descriptors by exploiting a protocol and implementation flaws in Tor and
derived onion addresses from them. The data collected allowed us to
analyse the landscape of Tor hidden services. We scanned the obtained
list for open ports, classified the content of Tor hidden
services that provide HTTP(S) service, and estimated the popularity of
obtained onion addresses.
We discovered a huge number of hidden services that are part of the ``Skynet''
botnet by looking for port number 55080. According to our analysis the
most popular services are HTTP, HTTPS, and SSH. 
We found that the content of Tor hidden services is rather varied.
The number of hidden services with illegal content or devoted to illegal activities
and the number of other hidden services (devoted to 
human rights, freedom of speech, anonymity, security, etc.) is almost the same;
among Tor hidden services one can even find a chess server.

Statistics of the popularity of hidden services look more distressing, however.
The most popular onion addresses are command and control centers of botnets
and resources serving adult content. The Silk Road market place is among
20 most popular hidden services.

In addition, we proposed a method for opportunistic deanonymisation of
hidden service clients and applied it to a popular hidden service.

Finally, we proposed a method to analyse the consensus history to detect tracking
of a hidden service. We applied it to ``Silkroad'' and were able to find 3 clear cases of tracking 
one of which was caused by our own experiments. In the second case an entity was
controlling 1 responsible HSDir but for a large timespan.
In the third case an entity was controlling all 6 responsible HSDirs
continuously for 24 hours.